\documentclass[namedreferences]{solarphysics}

\usepackage[hyperref,optionalrh,showbiblabels]{spr-sola-addons} 
\usepackage{graphicx}        
\usepackage{amssymb}        
\usepackage{color}           
\usepackage{breakurl}        

\usepackage{soul}

\chardef\us=`\_

\usepackage{siunitx}
\DeclareSIUnit\parsec{pc}

\begin{document}

\begin{article}
\begin{opening}

\title{\textsc{CMEchaser}, detecting line-of-sight occultations due to Coronal Mass Ejections}


\author[addressref=aff1,corref,email={shaifullah@astron.nl}]{\inits{G.~M.}\fnm{Golam}~\lnm{Shaifullah}~\orcid{0000-0002-8452-4834}}
\author[addressref=aff1,corref,email={tiburzi@astron.nl}]{\inits{C.}\fnm{Caterina}~\lnm{Tiburzi}~\orcid{0000-0001-6651-4811}}
\author[addressref=aff1,email={e-mail.c@mail.com}]{\inits{P.}\fnm{Pietro}~\lnm{Zucca}~\orcid{0000-0002-6760-797X}}
\address[id=aff1]{ASTRON, the Netherlands Institute for Radio Astronomy, Oude Hoogeveensedijk 4, Dwingeloo 7991PD, the Netherlands}

\runningauthor{Shaifullah, Tiburzi \& Zucca}
\runningtitle{Searching for occultations by Coronal Mass Ejections}

\begin{abstract}
We present a python-based tool to detect the occultation of
back-ground sources by foreground Solar coronal mass ejections. The
tool takes as input standard celestial coordinates of the source and translates
those to the Helioprojective plane, and is thus well suited for use with a wide variety
of background astronomical sources. This tool provides an easy means to search
through a large archival dataset for such crossings and relies on the
well-tested \texttt{Astropy} and \texttt{Sunpy} modules.
\end{abstract}
\end{opening}

\section{Introduction}
\label{sec:intro}
Magnetic activity of the Sun is responsible for the generation of solar eruptions such as Coronal Mass Ejections (CMEs) and solar flares \citep[see, e.g.,][]{Vrsnak2016}. Solar flares result in broadband electromagnetic emission, while CMEs are a  significant release of plasma with an embedded magnetic field \citep{Kahler1992}. CMEs directed towards the Earth may interact with the Earth's magnetosphere, leading to a number of well identified phenomena such as aurorae, magnetic reconnections on both the day and night sides of the Earth's magnetosphere and potentially adverse space weather conditions. A key parameter to forecast the impact of CMEs on the Earth's magnetosphere is its magnetic field direction and strength \citep[see, e.g.,][]{Vourlidas2000}.

Estimating the magnetic fields of CMEs is a challenging task as coronal densities are not suitable for direct estimation of the magnetic field by Zeeman effect, as is commonly done for the solar photosphere \citep{Howard1962}. The only means to probe a CME's magnetic field is by studying the Faraday rotation that it induces on the linearly-polarized radio emission of background sources, such as pulsars or quasars, during an occultation or via satellite borne probes, if those are fortunately located. In order to properly select the relative background sources, the location of the erupting CME needs to be well known in advance and this is to date still not easy and it represents a challenge.
For this reason, it is important to identify any astronomical background source with a significant linear polarization along the trajectory of the erupting CME.

In this article we present \textsc{CMEchaser}, a software able to calculate whether, during an observation, a certain astronomical object has been occulted by a CME. In \autoref{sec:sw} we thoroughly describe the software and its outputs. In \autoref{sec:demo} we demonstrate the effectiveness of the algorithm, while in \autoref{sec:applications} we show its possible applications. The conclusions are presented in \autoref{sec:conclusions}.

\section{A python-based searching algorithm for CME occultations}\label{sec:sw}
\textsc{CMEchaser} searches for the occultation of a background source by a CME given the observing epoch, the galactic coordinates of the source itself and the CME characteristics. While the original background sources utilised were radio pulsars, the software itself is agnostic of the details and can detect line-of-sight crossings for any given coordinate pair on the sky\footnote{See the Appendix for practical notes on the software requirements and usage}.
 \begin{figure}
 \begin{center}
 \includegraphics[width=0.52\columnwidth,keepaspectratio]{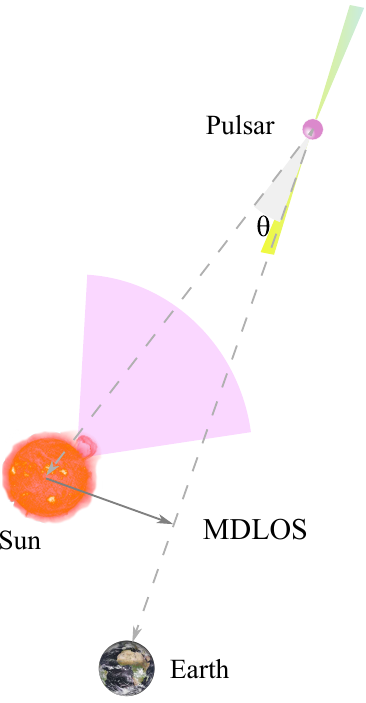}
 \caption{Sketch showing the geometric layout of the problem. The Solar (elongation) angle $\theta$ is the angle shaded in grey. MDLOS is the perpendicular distance from the Sun to the LoS to the pulsar from Earth, shown by the solid grey arrow. The pink shaded sector marks the cone of propagation of a CME, lifting of from a position angle of $\sim$\ang{345}, relative to the Solar North. Images of the Sun and the Earth were taken from publicly available ESA archives.}\label{fig:geometry}
 \end{center}
\end{figure}
\subsection{Code description}
The script requests as inputs the Galactic coordinates of the background source
along with the reference epoch at which they were determined and, if measured,
the background source's proper motion. The coordinates are then propagated to
the epoch of observation $T_0$ (note that this affects significantly only nearby
sources with very high proper motions). This is implemented via the
\texttt{astropy} library's SkyCoord class, which we use to store and
subsequently update the coordinates and proper velocity. The updated coordinates
are then transformed to the helioprojective plane, using the Stonyhurst
Stereographic projection \citep[see e.g.,][]{thompson06}, since that is the coordinate system used by the
SOHO/LASCO white light observations which we use to \textit{confirm} our
detections. This transformation follows conventions such as can be found in
\citet{meeus98}. Briefly, this is the same problem as measuring distances on a
geodesic and hence the techniques used are derived from pre-existing methods in
cartography. Specifically, we implemented the estimation of the helioprojective 
coordinates using the \citet{vincenty75} method. The Vincenty method can fail
for nearly antipodal points, and hence a fallback method using the
Haversine expansion is also provided. An excellent description of both the
translations can be found in \citet{skokic19}.

A correction to account for the misalignment between the ecliptic North and the
Solar north is also implemented at this stage, utilising methods from the
\texttt{sunpy} library. The angle theta between the Earth, the background source
and the Sun (see \autoref{fig:geometry}) is converted into the minimum (i.e.,
perpendicular) distance to the line-of-sight (MDLOS) from the Sun. The software
then selects all the CME events which, given a linear speed and associated
acceleration, will arrive at the line of sight at $T_0$.

For this, the software uses the SOHO/LASCO CME
catalogue\footnote{\url{https://cdaw.gsfc.nasa.gov/CME_list/}}. Note that this
catalogue is manually maintained and cannot be used for near real-time
detections. A more optimal catalogue for that use case would be an automated
one, such as the CACTus catalogue\footnote{\url{http://sidc.oma.be/cactus/}}.
However the CACTus catalogue is known to contain a number of false positive
detections of CMEs, and hence we prefer to implement the usage of the SOHO/LASCO
list as the default option.

For CMEs occurring in the mentioned time range, the code checks if the
background source, at the MDLOS position, falls into the sector into which the
CME is predicted to propagate given its position angle (relative to Solar North)
and opening angle as reported by the SOHO/LASCO catalog.
While the modelling of CME propagation is an involved exercise requiring very careful treatment, we assume that:
\begin{itemize}
    \item the CMEs from the SOHO/LASCO catalog have an average speed which is equivalent to the linear speed of the CME front, as measured from the LASCO white-light difference images;
    \item  the CME acceleration is constant;
    \item the angular width of the CME is constant.
\end{itemize}

Further, the software does not distinguish CMEs on the basis of their launch velocities relative to that of the Solar Wind, which is another complex problem that must often be modelled on a case by case basis to account for the dynamical properties of the Solar Wind, as well as the CME itself.

However, we point out that the objective of this code is to provide a means for identifying \textit{possible} crossings of CMEs in background source observations, and the actual detection and analysis will be carried out in forthcoming work on large collections of radio pulsar observations.

Among the outputs that will be described in Section \ref{sec:output}, \textsc{CMEchaser} produces a `quad' plot of the helioprojective plane, showing all of the CMEs that would have crossed the LoS at $T_0$. Then, for each of the CMEs identified above, the software downloads the nearest SOHO/LASCO and SDO white light images using the \texttt{Sunpy} interface to the Helioviewer webpage\footnote{\url{https://helioviewer.org/}} on which it overplots the position of the background source.

\subsection{Inputs}

The code expects two kinds of inputs. The first are the position of the background source in Galactic coordinates along with, where available and required, the proper motion terms. The second is a text-based archive of known CMEs, containing columns with headers identifying the launch angle (or position angle, PA) in the helioprojective plane, the linear speed and acceleration of the CME and the date and time of the CME launch. Currently the code is able to check for the presence of the SOHO/LASCO catalog and download it if necessary. However this can easily be substituted by a file containing columnar data in the same format of such catalog, which can be supplied by a command-line handle.

\subsection{Outputs}\label{sec:output}

\begin{figure}[!htbp]    
  \centerline{\hspace*{0.015\textwidth}
               \includegraphics[width=0.52\textwidth,clip=]{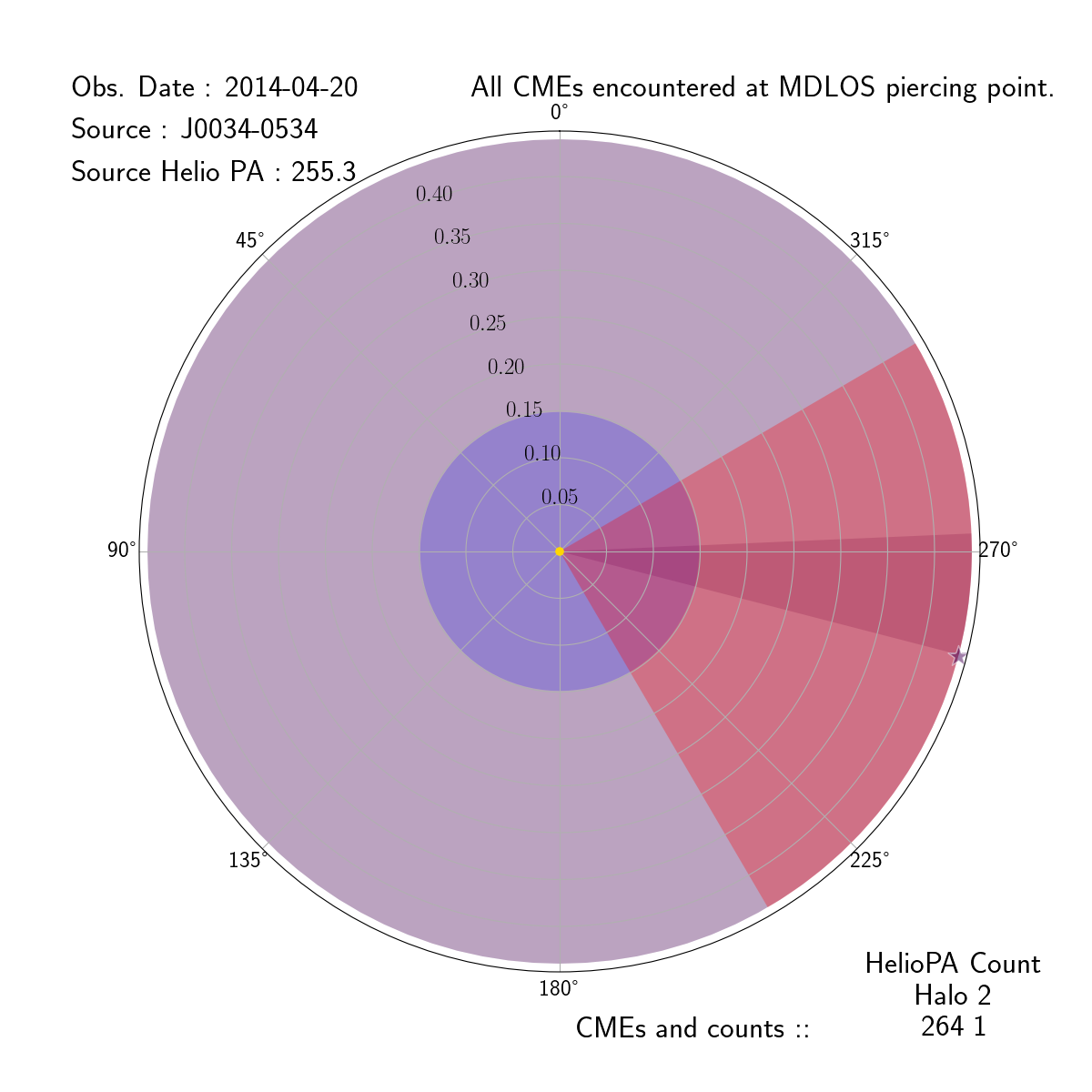}
               \hspace*{-0.03\textwidth}
               \includegraphics[width=0.515\textwidth,clip=]{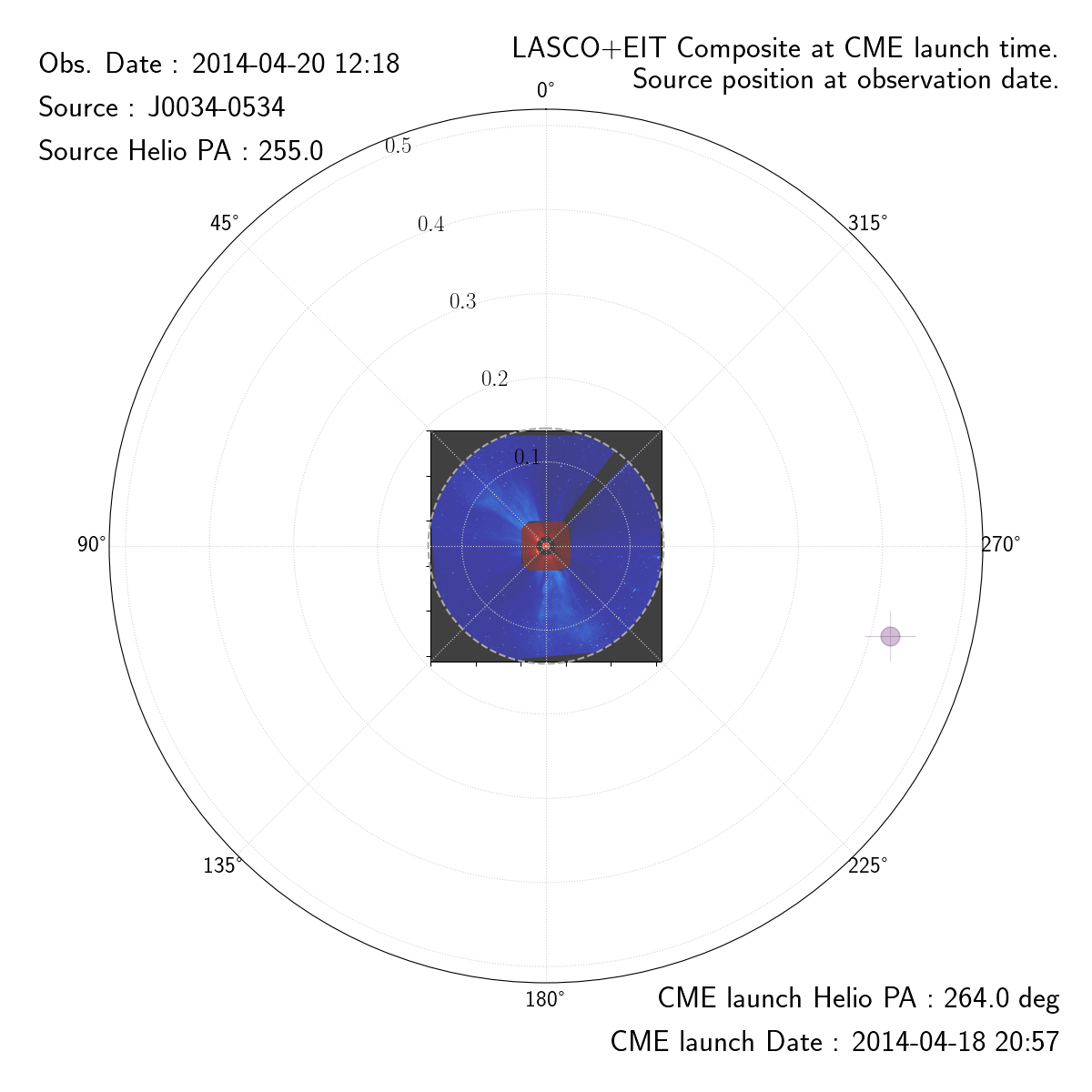}
             }
  \vspace{-1\baselineskip}   
  \centerline{\large \bf     
      \hspace{0.0 \textwidth} (a)
      \hspace{0.415\textwidth} (b)
         \hfill}
     \vspace{1\baselineskip}    
              
     \caption{Examples of the \textit{Quad} and \textit{C3} plots.
       \textbf{(a)}~\textit{Quad} plot, showing the detection of an
       occultation of the line of sight to the pulsar J0034$-$0534,
       shown by the five pointed star symbol. The pulsar was observed
       on the 20\textsuperscript{th} of April, 2014 at and was
       occulted by a halo CME launched on the 18\textsuperscript{th}.
       In this plot the CME is shown by the light green sector
       spanning the angular width of the CME (\ang{360} in this case)
       and the propagation is only drawn up to the Solar Elongation of
       the background source. The LASCO C3 field of view is shown by
       the purple shaded circle and the gold circle in the centre of
       the plot denotes the Sun. The radial axis shows the Solar
       Elongation converted to a projected separation in fractional
       \si{\astronomicalunit}.\label{fig:quad}
       \textbf{(b)}~An example of the \textit{C3} plot using the
       pulsar J0034$-$0534 as a background source, showing white light
       images from the SOHO and SDO timestamp closest to the CME
       launch time. In this plot, a CME which launched on the
       18\textsuperscript{th} of April, 2014 at 13:45 UTC is shown
       along with the position of the pulsar, which was observed on
       the 20\textsuperscript{th}. The LASCO C3 field of view is shown
       by the blue tinted region. The red square section of the image
       shows the LASCO C2 field of view while the centre of the image
       shows the SDO-AIA view of the Sun. The position of the pulsar
       is marked by the filled crosshair symbol at \ang{255}.\label{fig:c3}}%
   \label{fig:4panels}
   \end{figure}

\subsubsection{Plot outputs}
\textsc{CMEchaser} produces two kinds of plots: the `detection' plot showing all of the CMEs that might have crossed the LoS of the background source, and the `verification' plot, where for each CME in the detection plot, we produce an overlay of the \textit{nearest} SOHO/LASCO and SDO images with the background source. \\

The plots are named using the following convention:\\
$\langle{SourceName}\rangle$\_$\langle{Type}\rangle$\_$\langle{Date}\rangle$\_$\langle{SrcHelioPA}\rangle$\_$\langle{mdlos}\rangle$ where:
    \begin{itemize}
        \item SourceName: Name of background source supplied to script with -P
        \item Type: The kind of plot as described below:
        \begin{itemize}
        \item \textit{Quad}: Polar plot of all CMEs that will catch up with the
          LOS at $T_0$ (see text below)
        \item \textit{C3}: An overlay with the C3/LASCO image \textit{for each CME} that is on the Quad plot (see text below).
        \end{itemize}
        \item Date: For the quad plot, this is the date of observation of the \emph{source} while it is the date on which the \emph{CME} launched for the C3 plots.
        \item SrcHelioPA: The angle the LOS makes with the Solar North
        \item mdlos: The minimum distance to the line of sight in kpc
    \end{itemize}

The \emph{Quad} plot (\autoref{fig:4panels}a) is a polar plot of the source in the helioprojective plane, with coloured circles showing, in gold the Solar disc and in blue the size of the LASCO C3 field of view. A quadrant in red shows a \SI{90}{\degree} region around the heliographic position angle of the LoS.

Finally, we mark the individual CMEs that would have crossed the LoS. These are represented as sectors of widths equal to the CME width at launch and the propagation vector passing through the heliographic position angle of the CME;

The \emph{C3} plot in \autoref{fig:4panels}b is an overlay of the SDO image of the Sun at the centre and the LASCO/C3 image of each CME. The image used is the one \emph{at or closest to} the launch time. Along with these images, we plot the calculated position of the source in the helioprojective plane with the filled crosshair symbol.

\subsubsection{Text outputs}
Apart from the plots described above, we produce two sets of text outputs, one useful for any background source, and one specific for pulsars, where we predict the change in dispersion measures due to these CMEs. These are:
\begin{itemize}
    \item $\langle{SourceName}\rangle$\_FinalCMElist.txt contains the list of
      CMEs which have reached the LoS for the given pulsar observation date, extracted from the CDAW catalogue along with some extra columns:
    \begin{itemize}
        \item \textit{mdlos}: the minimum distance from the line of light (mdlos) for the given source which the CME crossed;
        \item \textit{Propagation\_time\_in\_s}: propagation time to mdlos, estimated using the linear speed and acceleration of the CME as provided by the CDAW catalogue;
        \item \textit{ArrivalTime}: Arrival time in MJD of the CME at the LOS.
    \end{itemize}
    \item $\langle{SourceName}\rangle$\_dmcme\_per\_obs.txt: a list of the excess DM due to the individual CMEs, sorted by observing date in MJD;
    \item $\langle{SourceName}\rangle$\_summed\_dmcme\_per\_obs.txt: total DM excess for each observing date in MJD. 
\end{itemize}
For the final two outputs, we currently utilise a simple spherical expansion model for the CME. Although this model is a relatively poor one for the propagation of CMEs in the Solar Wind, it allows us to quickly filter CMEs where we might be able to measure DM variations.

\section{Code demonstration}
\label{sec:demo}
 \begin{figure}
    \centering
    \includegraphics[width=.85\textwidth]{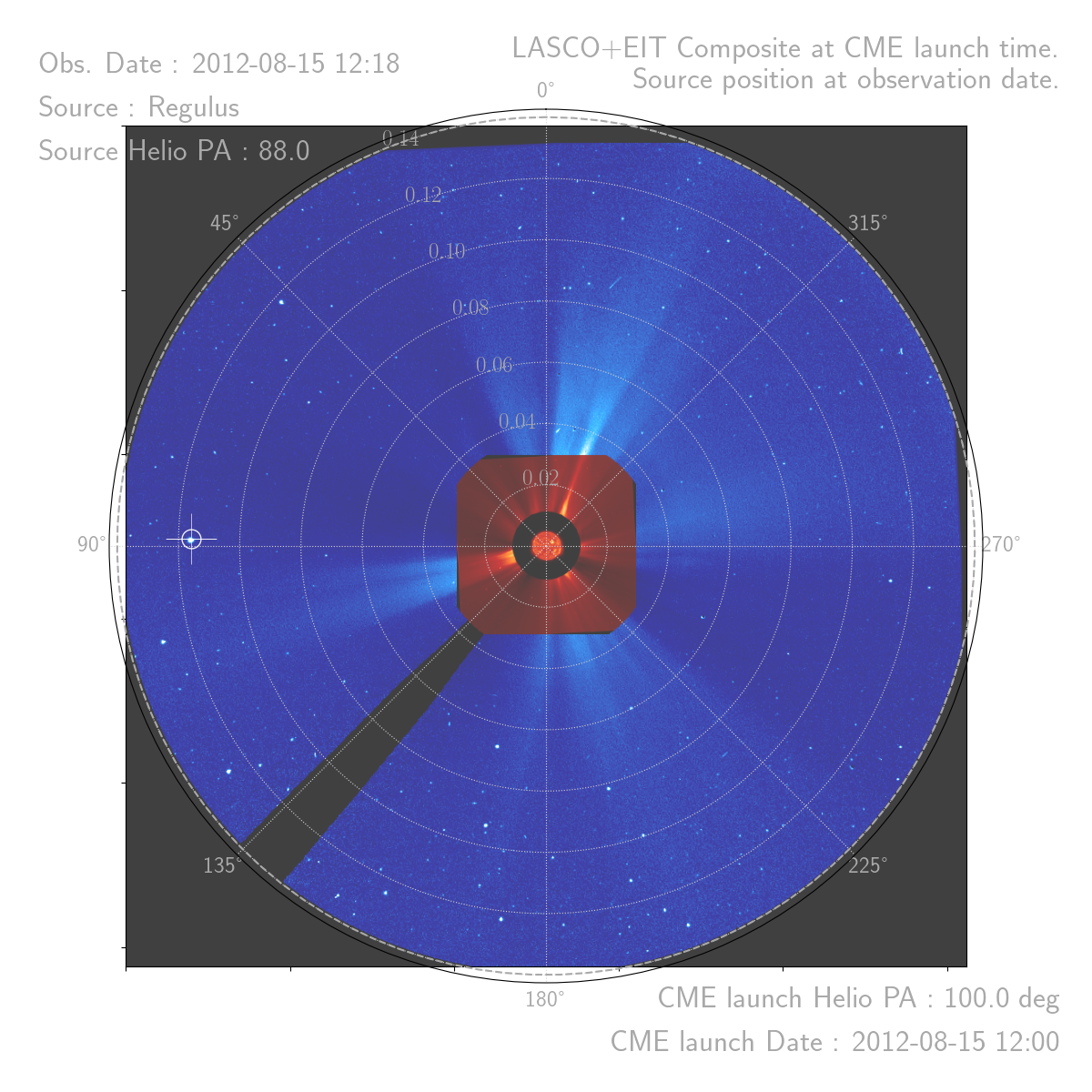}
    \vspace{-\baselineskip}
    \caption{\textit{C3} plot showing the position of Regulus at $\sim$12:00 UTC
      on the 15\textsuperscript{th} of August, 2012. On this date the background
      source is visible within the LASCO C3 field of view in an image taken at
      UTC 12:00. The point where the line of sight from the source to the Earth
      crosses the Heliocentric plane at UTC 12:18 is overplotted with a white
      crosshair symbol.}\label{fig:regulus}
\end{figure}

\begin{figure}
    \centering
    \includegraphics[width=.85\textwidth]{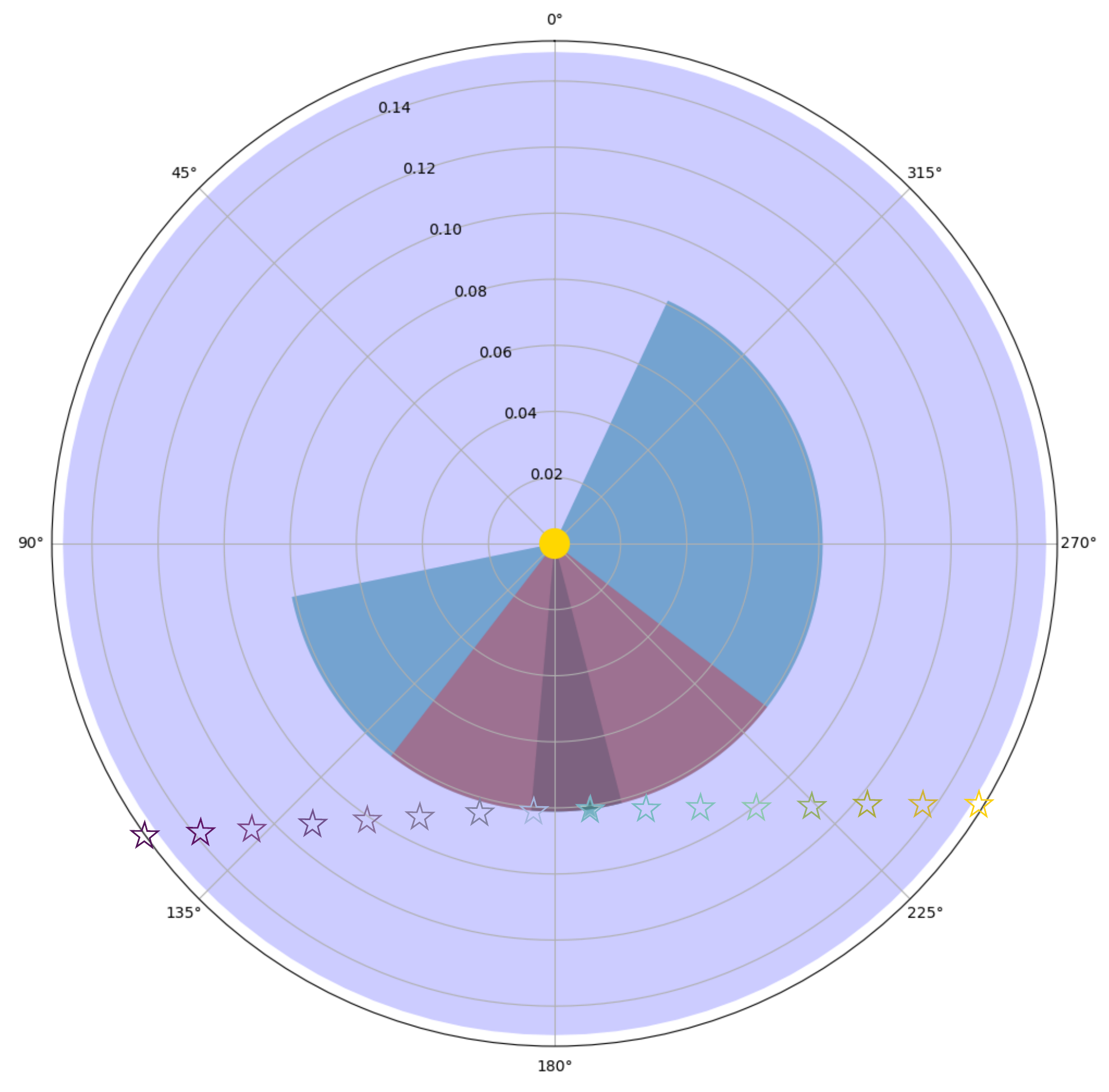}
    \caption{\textit{Quad} plot showing the position of PSR~B0950+08 tracked
      over the 13\textsuperscript{th} to the 28\textsuperscript{th} of August,
      2015, as viewed from Earth. Also shown are the tracks of two CMEs marked
      by blue-green shaded sectors which would have occulted the pulsar on the
      21\textsuperscript{st} of August. The filled green star symbols
      immediately to the right of the \ang{180} line shows the position of the
      pulsar on that date. The positions of the pulsar on other dates were
      overplotted manually to produce this composite image. The plot was only
      compared visually to the \protect\citet{hsd16} result, although
      orientations and positions were checked against both Helioviewer and
      common sky chart tools such as \href{https://stellarium.org/}{stellarium}.}
    \label{fig:howard}
\end{figure}

We now demonstrate the effectiveness of the software in correctly identify
astronomical objects in the field-of-view (FoV) of SOHO/LASCO (and SDO). For
this test, we refer to the graphical information available from the Sungrazer
project\footnote{\url{https://sungrazer.nrl.navy.mil/}}, where the position of
several known astronomical sources is identified during the Solar transit within
the SOHO/LASCO-C3 FoV. In particular, we focus on the position of the B-class
star Regulus on the 15th of August 2012 at 12:00 UTC. In \autoref{fig:regulus},
we show the observed position of Regulus, marked by the filled crosshair symbols
overlaid on the LASCO C3 composite. The optical position of the star is
marginally offset due to the time difference between the C3 image and the
observation epoch.
 
 Similarly, in \autoref{fig:howard}, we show the detected position of the pulsar, B0950+08 tracked from the 13\textsuperscript{th} to the 28\textsuperscript{th} of August, 2015. The track follows that observed by \citet{hsd16}, where the authors observe a CME crossing the pulsar on the 21\textsuperscript{st} of August.  In the example plot shown here, two CMEs which were launched from heliospheric position angles of \ang{255} and \ang{148} on the 20\textsuperscript{th} of August are detected to have crossed the LoS to the pulsar. The pulsar was assumed here to have been observed at 12:00 UTC and the CMEs are marked as blue-green shaded sectors, whose widths are equal to the \ang{160} and \ang{93}, respectively.
 
 

\section{Applications}
\label{sec:applications}
The most immediate application of \textsc{CMEchaser} is for mining archival datasets of any kind of polarized astronomical sources (e.g., long-term pulsar monitoring such the ones described in \citealt{tvs19}). This would allow to select the observations allegedly affected by the passage of a CME and study their properties with respect of the baseline offered by the rest of the dataset.\\

Besides this, though, \textsc{CMEchaser} can be easily adapted to be utilised to identify linearly polarized background sources when a CME is launched, and at least its PA and velocity are identified. In fact, as mentioned earlier, the presented software can accept customized CME catalogs in input $-$ thus, if a CME occur it will be possible to create a catalog with the same format as of the SOHO/LASCO one and parse it to \textsc{CMEchaser} to select linearly polarized sources whose LoS will be occulted. This application is useful in case of  \textit{target-of-opportunity} (ToO) observing proposals dedicated to the scientific study of CMEs, but also to trigger warnings in the context of space weather monitoring. As mentioned in \autoref{sec:intro}, the recombination of a CME's and the Earth's magnetic fields allows to charged particles carried from the CME itself to stream down the field lines of the geomagnetic field, compromising the activity of satellites, power grids, geolocalization systems etc. The presented software can be used in forecasting tactics to understand the orientation of the magnetic field of an incoming CME and the risk level associated with it. As a matter of fact, it has to be noticed that while it is known that the launch of a CME is likely during periods of high Solar activity, a timely prediction of a CME occurrence is currently impossible. Further, the dynamics of CME evolution is a region of active research and experimental verification of the expected behaviour of CMEs far away from the Sun are valuable inputs not only to simulation tools but also to forecasting efforts.\\

\section{Conclusions}
\label{sec:conclusions}
We have presented \textsc{CMEchaser}, a software able to show if an astronomical object has been occulted by a CME within two days from the observation of such object. For this scope, \textsc{CMEchaser} takes in input the object position and the observation date and searches for suitable CME events in the SOHO/LASCO CME catalogue, giving in output text files containing information about the CME that has possibly occulted the astronomical source and plots of the field-of-view of the LASCO cameras overlaid with the position of the object at the time in which the observation was performed. 

\textsc{CMEchaser} has applications in space weather studies, as it will increase the regularity with which it is possible to detect a CME occultation is information about the CME (at least its position angle and emission epoch) are provided. For this exercise to be successful, it is necessary to continue ToO programs combined with trigger observations of, e.g., Solar activity in K-band \footnote{See, e.g., the \textsc{Sundish} project (\url{https://sites.google.com/inaf.it/sundish/}).}, or Solar bursts.

\begin{acks}
GS is supported by the Netherlands Organisation for Scientific Research NWO (TOP2.614.001.602; P.I. - Dr. G. Janssen). CT is supported by a VENI fellowship (016.Veni.192.086) awarded by the Dutch Research Council (NWO), and this work is part of the research programme SOLTRACK.

This research utilises  Sunpy (version $>= 1.0.0$)\footnote{https://sunpy.org,
  \protect\href{https://doi.org/10.5281/zenodo.3096966}{DOI:
    10.5281/zenodo.3096966}} and Astropy (version $>= 3.0.5$)\footnote{http://www.astropy.org, \protect\href{https://doi.org/10.5281/zenodo.2556700}{DOI: 10.5281/zenodo.2556700}}  community-developed core Python packages for Solar \citep{sunpy_community2020} and General Astronomy \citep{apy13,apy18}, respectively. 
\end{acks}



\appendix   
 \section{Practical notes}\label{appendix}
 \textsc{CMEchaser} can be downloaded from \url{https://bitbucket.org/golamshaifullah/cme_chaser/wiki/Home}.\\ 
 The software is a fully python-based code and to be run, it needs the most common python packages (numpy, scipy etc), along with the \textsc{Sunpy} and \textsc{Astropy} modules. We advise running it after having created and activated the \textsc{CONDA} environment as shown on the \textsc{CMEchaser} wiki to avoid issues with missing packages and libraries.\\
 The software can be run in two ways, depending on the background source that the user wants to search the occultation for. If the background object is a pulsar present in the \textsc{psrcat} software\footnote{https://www.atnf.csiro.au/research/pulsar/psrcat/}, the user can simply pass the name of the source as referred to the J2000 epoch (e.g., J1022+1001). Alternatively, if the background object is a pulsar missing in the catalogue, or any other source, the user can run the software by supplying the source name, its Galactic latitude and longitude, proper motion in Galactic coordinates and the reference epoch for the astrometric parameters.\\
 To test whether the setting up of \textsc{CMEchaser} was successful, we
 recommend trying one of the example commands in
 \url{https://bitbucket.org/golamshaifullah/cme_chaser/wiki/commands_for_star_check.md}.
 These will lead to the identification of position of a known star (i.e.,
 visible in the optical) against the Helioviewer database through the use of the
 customary, fake catalog \textsc{myuniv.txt}. Note that, in a real run, the user
 can supply their own customised CME catalog as a simple text or
 comma-separated-values file with the column header for the CME launch date set
 to \texttt{dme}, or not parse any catalog. In this last case,
 \textsc{CMEchaser} will automatically download and use the SOHO/LASCO catalog.
 

\bibliographystyle{spr-mp-sola}
\bibliography{cmechaser}

\IfFileExists{\jobname.bbl}{} {\typeout{}
\typeout{****************************************************}
\typeout{****************************************************}
\typeout{** Please run "bibtex \jobname" to obtain} \typeout{**
the bibliography and then re-run LaTeX} \typeout{** twice to fix
the references !}
\typeout{****************************************************}
\typeout{****************************************************}
\typeout{}}

\end{article} 

\end{document}